\documentstyle[12pt]{article}
\input amssym.def
\input amssym
\def \dst {\displaystyle}
\def \txt {\textstyle}

\def \circs{{{\viipt\circ}\atop{\viipt\circ}}}

\date{\centerline{3/30/95}}
\title{Quantum Sturm-Liouville Equation,\\
Quantum Resolvent, Quantum Integrals,\\
and Quantum KdV : the Fast Decrease Case.}
\author{M. Zyskin\thanks{Department of Physics and Astronomy, Rutgers
University, Piscataway, NJ 08855}}

\begin{document}
\maketitle
\begin{abstract}
We construct quantum operators solving the quantum versions of the
Sturm-Liouville equation and the resolvent equation, and show the
existence of conserved currents. The construction depends on the
following input data: the basic quantum field $O(k)$ and the
regularization .

\end{abstract}
\newpage

\section*{Introduction.}

To quantize nonlinear integrable systems, it was suggested by Faddeev
et al. to quantize first the scattering data [1,2,6]. The scattering data
are action-angle type fields, and their time dependence (in
the Heinseberg formalism)
for  the case of fast decrease is trivial. At fixed time, scattering
data are operators with some commutation relations. We will
show, how to construct from scattering operators the observable fields, namely,
the densities of the conserved currents. For simplicity,we consider
the fast decrease case and the solitonless sector only.In the solitonless
sector,the scattering data for
the clasical Sturm-Liouville equation are the transition coefficent $a(k)$
and the reflection coefficient $b(k)$,or their ratio $O(k)=b(k)/a(k)$.We
will assume,that in the quantum theory we are given the operator $O(k)$,
and we will proceed to define the observable fields in the theory.We will
see,that we do not need specially to have a quadratic algebra for $a$ and $b$
in order to have the conserved currents in the theory.

In order to build the conserved currents, we use the Nonlinear Fourier
Transform developed for the  classical integrable equations in the paper [8].
In that paper, the authors investigated the nonlinear nonlocal change of
variables,
given by certain nonlinear functionals , with kernels given by  homogeneous
generalized functions. This nonlinear change of variables reduces the issue
of integrability of the nonlinear equation to certain identities for rational
symmetric functions. (Such propery is similar to the usual Fourier Transform,
which reduces the linear partial differential equation with constant
coefficients to the polynomial algebraic equation).In the quantum theory,
we use the functionals of the same type as we had in the classical case to
express the conserved currents through 'scattering data' fields. Now
the 'scattering data' are quantum fields,so one have to smear them to
get rid of the singularities in the product, and give the prescription for the
ordering.The best thing to do is to use the symmetric (Weyl) order.With the
Weyl ordering, it is not nessessary to know the commutation relations for
the quantum field $ O(k)$ in order to construct the conserved currents;  in
fact, we can start from  any  $ O(k)$, and write the conserved currents
as functionals of $O(k)$. (Of course,the correlation functions of such
currents do depend on what $O(k)$ we have choosen , but the functionals,
expressing the currents through $O(k)$ , are universal.)

\newpage
\section*{The construction of the conserved currents.}
Let $O(k)$, ``the basic field'', be some reasonable quantum field,
defined as operator by matrix elements in a
Hilbert space,or through bosonization in the style of [6].
Here $k \in {\Bbb R}$ (we consider the solitonless case,
for the solitons one has to add finite number of points $\{ i\kappa_m
\}^M_{m = 1}, \kappa_m \in {\Bbb R}^+$).

Let $\{ \Delta_n(k) \}^{\infty}_{n = 1}$ be a sequence of
$C^{\infty}$ test functions of fast decrease which converge to the
$\delta$-function, say $\Delta_n(k) = \dst\frac{1}{2\sqrt{\pi n^2}}
\exp \left( -\dst\frac{|k|^2}{4n^2} \right)$, $n = 1, 2, \ldots$ and
$e_m(k)$ be a $C^{\infty}$ function which is equal to 1 for $|k| < m$
and of fast decrease for $|k| > 2m$, $m = 1, 2, \ldots$.

For $S = (s_1, s_2),\ s_1, s_2 = 1, 2, \ldots$ define the smeared
field
\setcounter{equation}{0}
\begin{equation}
O_S(k) = e_{s_1}(k) \int O(k_1) \Delta_{s_2}(k - k_1) dk_1.
\end{equation}

{}From the general axioms for quantum fields,and from the experience,
we expect the products of smeared fields to be smooth in $\{k\}$
(including the diagonals). Let
$$
(O_S(k_1) O_S(k_2) \ldots O_S(k_n))_W = \dst\sum_{\txt\{\dst\sigma \txt\}}
\frac{1}{n!}
O_S(k_{\sigma(1)}) O_S(k_{\sigma(2)}) \ldots O_S(k_{\sigma(n)})
$$
\noindent be the Weyl ordered product of smeared fields.

\vspace*{6pt}
{\bf Remarks}.
\begin{enumerate}
\item  Actually, we need only smooth behavior at points $k_i + k_j =
0$. If in the products $O(k_1) O(k_2) \ldots O(k_n)$ singularities
are not at $k_i + k_j = 0$ we can take $\Delta(k) = \delta(k)$, and
leave only the regularization by $e_S(k)$.

\item The only property that we need from the product of operators \newline
$ (O_S(k_1) O_S(k_2) \ldots O_S(k_n))_W $ is that it should be symmetric in
$ \{k\}$ .Therefore,if $ O(k) $ would be a free field, we could use
the  Wick order, rather than the Weyl order.

\end{enumerate}

Consider the following operator-valued series
\begin{equation}
\begin{array}{l}
\Psi_{S, \lambda}(k, x) = 1 + \dst\sum^{\infty}_{n = 1} \frac{\lambda^n}{n!}
\int (O_S(k_1) O_S(k_2) \ldots O_S(k_n))_W
\\ [8pt]
\dst\frac{1}{(k + k_1 + i0)(k_1 + k_2 + i0) \ldots (k_{n - 1} + k_n +
i0)}
\\ [8pt]
\cdot \exp(2i(k_1 + \ldots + k_n) x) \dst\frac{dk_1 \ldots
dk_n}{(2\pi i)^{n}}
\end{array}
\end{equation}

\begin{equation}
\begin{array}{l}
i_{S, \lambda}(0, x) \equiv - \dst\frac{1}{2}

\\ [18pt]

i_{S, \lambda}(p, x) = \dst\sum^{\infty}_{n = 1} \frac{\lambda^n}{n!} \int
(O_S(k_1) O_S(k_2) \ldots O_S(k_n))_W
\\ [12pt]
\dst\frac{k_1^{2p - 1} + k_2^{2p - 1} + \ldots + k_n^{2p -
1}}{(k_1 + k_2 + i0)(k_2 + k_3 + i0) \ldots (k_{n - 1} + k_n + i0)}
\\ [12pt]
\exp( 2i(k_1 + \ldots + k_n)x) \dst\frac{dk_1 \ldots dk_n}{(2\pi i)^{n}},
\hskip65pt p = 1, 2, \ldots

\\ [30pt]

u_{S, \lambda}(x) = 4i_{S, \lambda}(1, x)
\end{array}
\end{equation}
\vspace*{6pt}

\noindent Each term in the series is well-defined, because $(O_S(k_1)
O_S(k_2) \ldots O_S(k_n))_W$ is smooth in $k_1, k_2, \ldots, k_n$ and of fast
decrease, and $\dst\frac{1}{\dst\sqcap^{n - 1}_{j = 1} (k_j + k_{j + 1} +
i0)}$ is a distribution in $n$ variables.

\vspace*{6pt}

In the classical limit, $O(k)$ is some function (scattering data),and
\begin{equation}
\begin{array}{l}
\Psi(k, x) = \dst\int^{\infty}_0 e^{-\lambda} \Psi_{\lambda}(k, x)
d\lambda
\\ [12pt]
i(p, x) = \dst\int^{\infty}_0 e^{-\lambda} i_{\lambda}(p, x) d\lambda,
\hskip30pt p = 1, 2, \ldots
\end{array}
\end{equation}

\noindent are ,respectively,the Jost function and the densities of integrals of
motion, expressed through scattering data ,which were considered in [8].

In the quantum case, the integral over $\lambda$ should be
regularized to get the convergent answer.For our purposes,
the precise prescription to take the final integral over $\lambda$ is
not important.
\vskip12pt

{\bf  Remark.}
We assumed that the matrix elements of integrals in (2) , (3) grow with n not
faster than (n-1)! for large $n$ ,and therefore,the series (2),(3) are
convergent ; the operarators (4) are related to operators (2), (3) via
the (regularized) Borel transform. In fact,
if we take some particular $ O(k)$ ,we may need the generalized Borel transform
to ensure convergence; namely, we have to evaluate the large $n$ dependence of
the integrals and to change the coefficient
in front of the n-th term integral from $\frac{\lambda^n}{n!}$ to some
$c(n,\lambda)$ in all the formulas in such a way that the series,with the
coefficients $c(n, \lambda)$ instead of $\frac{\lambda^n}{n!}$,are convergent.
Afterwards,we introduce the the measure  $ d\mu(\lambda)$ ;the regularized
operators $\Psi_S(k,x)$ and $i_S(p,x)$ are defined as
$$
\begin{array}{l}
\Psi_S(k, x) = \dst\int d\mu(\lambda) \Psi_{S,\lambda}(k, x)
\\ [12pt]
i_S(p, x) = \dst\int d\mu(\lambda)  i_{S,\lambda}(p, x) ,
\hskip30pt p = 1, 2, \ldots
\end{array}
$$

\vskip12pt

\noindent{\bf Definition}.

\noindent Let us have 2 quantum fields,
$$
A_{S,\lambda}(x) := \sum_n \frac{\lambda^n}{n!} \int (O_S(k_1) O_S(k_2) \ldots
O_S(k_n))_W\
f_n(k_1, k_2, \ldots, k_n, x)\ \frac{dk_1 \ldots dk_n}{(2\pi i)^n}
$$
$$
B_{S,\lambda}(x) := \sum_m \frac{\lambda^m}{m!} \int (O_S(k_1) O_S(k_2) \ldots
O_S(k_m))_W\
g_n(k_1, k_2, \ldots, k_m, x)\ \frac{dk_1 \ldots dk_m}{(2\pi i)^n}
$$
where $\{ f_n \},\ \{ g_m \}$ are some distributions.

Define the ${\circs\circs}$ product of $A_{S,\lambda}(x)$ and
$B_{S,\lambda}(x)$ as follows:
\begin{equation}
\begin{array}{l}
\circs A_{S,\lambda}(x) B_{S,\lambda}(x)\circs = \dst\sum_n \sum_{m_1, m_2\ \ \
\atop m_1 + m_2 = n} \frac{\lambda^n}{n!} \int (O_S(k_1) O_S(k_2)
\ldots O_S(k_n))_W
\\ [8pt]
f_{m_1} (k_1, k_2, \ldots, k_{m_1}, x)\ g_{m_2} (k_{m_1 + 1}, k_{m_1 +
2}, \ldots, k_n, x)\ \dst\frac{dk_1 dk_2 \ldots dk_n}{(2 \pi i)^n}
\end{array}
\end{equation}

{\bf Remark.}

Suppose that the integrals
$$
\begin{array}{l}
A_{S}(x) :=\dst\int^{\infty}_0 e^{-\lambda} A_{S,\lambda}(x) d\lambda =
\\ [8pt]
\dst\sum_n \dst\int (O_S(k_1) O_S(k_2) \ldots O_S(k_n))_W\
f_n(k_1, k_2, \ldots, k_n, x)\ \frac{dk_1 \ldots dk_n}{(2\pi i)^n}
\end{array}
$$
$$
\begin{array}{l}
B_{S}(x) :=\dst\int^{\infty}_0 e^{-\lambda} B_{S,\lambda}(x) d\lambda =
\\ [8pt]
\dst\sum_m \dst\int (O_S(k_1) O_S(k_2) \ldots O_S(k_m))_W\
g_m(k_1, k_2, \ldots, k_m, x)\ \frac{dk_1 \ldots dk_m}{(2\pi i)^m}
\end{array}
$$
\vspace*{14pt}
$$
\begin{array}{l}
\circs A_{S}(x)B_{S}(x) \circs := \dst\int^{\infty}_0 e^{-\lambda}
\circs A_{S,\lambda}(x) B_{S,\lambda}(x)\circs d\lambda =
\\ [8pt]
\dst\sum_n \sum_{ \ m_1, m_2\
\atop m_1 + m_2 = n} \int (O_S(k_1) O_S(k_2)
\ldots O_S(k_n))_W
\\ [8pt]
f_{m_1} (k_1, k_2, \ldots, k_{m_1}, x)\ g_{m_2} (k_{m_1 + 1}, k_{m_1 +
2}, \ldots, k_n, x)\ \dst\frac{dk_1 dk_2 \ldots dk_n}{(2 \pi i)^n}
\end{array}
$$
are well-defined and absolutely convergent (in the sence that all the matrix
elements of these operators are absolutely convergent ). If,furthermore,
$O(k)$ were commuting operators, the symmetrization $(O_S(k_1) O_S(k_2) \ldots
O_S(k_n))_W$  would not give anything , and we would have $A_{S}(x) \cdot
B_{S}(x) =
\circs A_{S}(x)B_{S}(x) \circs$. Actually, in the quantum case the operators
$O(k)$ do not commute. Therefore, the
products $A_{S,}(x)\cdot B_{S,}(x)$ \quad and
$\circs A_{S}(x)B_{S}(x)\circs $ are different; the
difference is an infinite series in the commutators of the operators
$O(k)$ .
\vspace*{14pt}

\noindent{\bf Lemma 1} (``Quantum Sturm-Liouville equation'').

Operator $\Psi_{S, \lambda}(k, x)$ satisfies the
following equation:
$$
\left( \dst\frac{d^2}{dx^2} + 2ik \frac{d}{dx}
\right) \Psi_{S, \lambda}(k, x) = - \circs u_{S, \lambda}(x) \Psi_{S,
\lambda}(k, x) \circs~,
$$
where
$$
\begin{array}{l}
\circs u_S(x) \Psi_{S, \lambda}(k, x)\circs = 4 \dst\sum^{\infty}_{n = 1}
\frac{\lambda^n}{n!} \int \left( O_S(k_1) O_S(k_2) \ldots O_S(k_n) \right)_w
\\ [12pt]
\left( \dst\frac{k_1 + k_2 + \ldots + k_n}{(k_1 + k_2 + i0)(k_2 + k_3
+ i0) \ldots (k_{n - 1} + k_n + i0)} \right.
\\ [12pt]
+ \dst\sum^{n - 1}_{m = 1} \frac{1}{(k + k_1 + i0)(k_1 + k_2 + i0)
\ldots (k_{m - 1} + k_m + i0)}
\\ [12pt]
\left. \dst\frac{(k_{m + 1} + k_{m + 2} + \ldots + k_n)}{(k_{m + 1} + k_{m + 2}
+ i0)(k_{m + 2} + k_{m + 3} + i0) \ldots (k_{n - 1} + k_n + i0)} \right)
\\ [12pt]
\exp(2i(k_1 + \ldots + k_n)x) \dst\frac{dk_1 dk_2 \ldots dk_n}{(2 \pi i)^n}
\end{array}
$$

\noindent{\bf Lemma 2} (``Quantum Resolvent Equation'').

Operators $i_{S, \lambda}(p, x)$ satisfy the following equation:
\begin{equation}
\begin{array}{l}
-4 \dst\frac{d}{dx} i_{S, \lambda}(p + 1, x) = \dst\frac{d^3}{dx^3}
i_{S, \lambda}(p, x) + 4 \txt\circs u_{S, \lambda}(x) \left( \dst\frac{d}{dx}
i_{S, \lambda}(p, x) \right)\txt\circs
\\ [8pt]
+ 2 \circs \left( \dst\frac{d}{dx} u_{S, \lambda}(x) \right) i_{S,
\lambda}(p, x)\circs\ \quad , p = 0, 1, 2, \ldots
\end{array}
\end{equation}

where for $p = 1, 2, \ldots$
$$
\begin{array}{l}
\circs u_{S, \lambda}(x) \left( \dst\frac{d}{dx} i_{S, \lambda}(p, x)
\right)\circs = 4 \dst\sum^{\infty}_{n = 2} \frac{\lambda^n}{n!}
\dst\int \left( O_S(k_1) O_S(k_2) \ldots O_S(k_n) \right)_W
\\ [12pt]
\dst\sum^{n - 1}_{m = 1} \dst\frac{(k_1 + k_2 + \ldots + k_m)}{(k_1 +
k_2 + i0)(k_2 + k_3 + i0) \ldots (k_{m - 1} + k_m + i0)}
\\ [12pt]
\cdot \dst\frac{(k^{2p - 1}_{m + 1} + k^{2p - 1}_{m + 2} + \ldots +
k^{2p - 1}_n)(k_{m + 1} + \ldots + k_n)}{(k_{m + 1} + k_{m + 2} +
i0)(k_{m + 2} + k_{m + 3} + i0) \ldots (k_{n - 1} + k_n + i0)}
\\ [12pt]
\exp(2i(k_1 + \ldots + k_n)x) \dst\frac{dk_1 \ldots dk_n}{(2 \pi i)^n};
\end{array}
$$
and
$$
\begin{array}{l}
\circs \left( \dst\frac{d}{dx} u_{S, \lambda}(x) \right) i_{S, \lambda}(p, x)
\circs = 4 \dst\sum^{\infty}_{n = 2} \frac{\lambda^n}{n!}
\dst\int \left( O_S(k_1) O_S(k_2) \ldots O_S(k_n) \right)_W
\\ [12pt]
\dst\sum^{n - 1}_{m = 1} \dst\frac{(k_1 + k_2 + \ldots + k_m)^2}{(k_1 +
k_2 + i0)(k_2 + k_3 + i0) \ldots (k_{m - 1} + k_m + i0)}
\\ [12pt]
\cdot \dst\frac{(k^{2p - 1}_{m + 1} + k^{2p - 1}_{m + 2} + \ldots +
k^{2p - 1}_n)}{(k_{m + 1} + k_{m + 2} + i0)(k_{m + 2} + k_{m + 3} + i0)
\ldots (k_{n - 1} + k_n + i0)}
\\ [12pt]
\exp (2i(k_1 + \ldots + k_n)x) \dst\frac{dk_1 \ldots dk_n}{(2 \pi i)^n}
\end{array}
$$

{\bf Lemma 3} (``Quantum Resolvent'').

We call the formal series
$$
R_{S, \lambda}(k, x) := 1 - 2 \dst\sum^{\infty}_{n = 1}\frac{1}{k^{2n}}
i_{S, \lambda}(n,x)
$$
\noindent the Quantum Resolvent of the Sturm-Liouville Equation.
Here $i_{S, \lambda}(n , x)$ are defined in (3) ; by Lemma 2, these operators
$i_{S, \lambda}(n , x)$ satisfy (6).

There is the following relation between $R$ and $\Psi$:

\begin{equation}
R_{S, \lambda}(k, x) = \circs \Psi_{S, \lambda}(k, x) \Psi_{S,
\lambda}(-k, x)\circs ,
\end{equation}

\noindent where,from the definition of the $ \circs \circs$
product (5)
$$
\begin{array}{l}
\circs \Psi_{S, \lambda}(k, x) \Psi_{S, \lambda}(-k, x) \circs = 1 +
\dst\sum^{\infty}_{n = 1}
\frac{\lambda^n}{n!} \int \left( O(k_1) O(k_2) \ldots O(k_n) \right)_W
\\ [12pt]
\left( \dst\frac{1}{(k + k_1 + i0)(k_1 + k_2 + i0) \ldots (k_{n - 1}
+ k_n + i0)} \right.
\\ [12pt]
+ \dst\frac{1}{(-k + k_1 + i0)(k_1 + k_2 + i0) \ldots (k_{n - 1} +
k_n + i0)}
\\ [12pt]
+ \dst\sum^{n - 1}_{m = 1} \frac{1}{(k_1 + k_2 + i0)(k_2 + k_3 + i0)
\ldots (k_m + k + i0)}
\end{array}
$$
$$
\begin{array}{l}
\left. \dst\frac{1}{(-k + k_{m + 1} + i0)(k_{m + 1} + k_{m + 2} + i0)
\ldots (k_{n - 1} + k_n + i0)} \right)
\\ [12pt]
\exp(2i(k_1 + \ldots + k_n)x) \dst\frac{dk_1 \ldots dk_n}{(2 \pi i)^n}
\end{array}
$$
The equation (7) should be understood in the following way : the right-hand
side of (7) depends from $k$ only through the denominator. We have to expand it
formally in the inverse powers of $k$ . The series we  obtain in this way
is $R_{S, \lambda}(k, x) $.

{\bf Lemma 4}

There is the following relation for the operators  $i_{S, \lambda}(n, x)$ :

$$
\begin{array}{l}
2 \circs i_{S, \lambda}(p_1, x) \dst\frac{d}{dx} i_{S, \lambda}(p_2,
x) \txt\circs - \dst\frac{d}{dx} \left( \txt\circs i_{S, \lambda}(p_1, x)
i_{S, \lambda}(p_2, x)\circs \right)
\\ [12pt]
:= 4i \dst\sum^{\infty}_{n = 2} \frac{\lambda^n}{n!} \int \left( O_S(k_1)
O_S(k_2) \ldots O_S(k_n) \right)_W
\\ [12pt]
\cdot \dst\sum^{n - 1}_{m = 1} \frac{k_1^{2p_1 - 1} + k_2^{2p_1 -
1} + \ldots + k_m^{2p_1 - 1}}{(k_1 + k_2 + i0)(k_2 + k_3 + i0)
\ldots (k_{m - 1} + k_m + i0)}
\\ [12pt]
\dst\frac{k_{m + 1}^{2p_2 - 1} + k_{m + 2}^{2p_2 -
1} + \ldots + k_n^{2p_2 - 1}}{(k_{m + 1} + k_{m + 2} + i0)(k_{m +
2} + k_{m + 3} + i0) \ldots (k_{n - 1} + k_n + i0)}
\\ [12pt]
(-k_1 - k_2 - k_3 - \ldots - k_m + k_{m + 1} + \ldots + k_n)
\\ [12pt]
\exp \left( 2i(k_1 + k_2 + \ldots + k_n)x \right)\dst\frac{dk_1 \ldots dk_n}{(2
\pi i)^n}
\\ [36pt]
= -2i \dst\sum^{\infty}_{n = 1} \frac{\lambda^n}{n!} \int \left( O_S(k_1)
O_S(k_2) \ldots O_S(k_n) \right)_W
\\ [12pt]
\cdot \Bigl( (k_1^{2p_1 + 1} + k_2^{2p_1 + 1} + \ldots +
k_n^{2p_1 + 1}) \cdot (k_1^{2p_2 - 1} + k_2^{2p_2 - 1} + \ldots +
k_n^{2p_2 - 1}) \hskip10pt -
\\ [12pt]
(k_1^{2p_1 - 1} + k_2^{2p_1 - 1} + \ldots +
k_n^{2p_1 - 1}) \cdot (k_1^{2p_2 + 1} + k_2^{2p_2 + 1} + \ldots +
k_n^{2p_2 + 1}) \hskip30pt \Bigr)
\\ [12pt]
\cdot \dst\frac{\exp (2i(k_1 + \ldots + k_n)x) }{(k_1 + k_2 + i0)(k_2 + k_3 +
i0) \ldots (k_{n - 1}+ k_n + i0)} \dst\frac{dk_1 \ldots dk_n}{(2 \pi i)^n}.
\end{array}
$$

\vspace*{20pt}
\newpage
The field $O_S(k)$ is of action-angle type, and its time evolution is
trivial
$$
-i \frac{\partial}{\partial t_{(l)}} O_S^{(l)} (k, t_{(l)}) = 2k^{2l
- 1} O_S^{(l)} (k, t_{(l)}),\ l = 2, 3, \ldots \quad .
$$

\vspace*{20pt}
{\bf Lemma 5} (``Quantum Integrals'').

The quantities
$$
\begin{array}{l}
i_{S, \lambda}(p_1, l, x, t_{(l)}) = \dst\sum^{\infty}_{n = 1}
\frac{\lambda^n}{n!}
\int \left( O_S(k_1) O_S(k_2) \ldots O_S(k_n) \right)_W
\\ [12pt]
\dst\frac{k_1^{2p_1 - 1} + k_2^{2p_1 - 1} + \ldots + k_n^{2p_1
- 1}}{(k_1 + k_2 + i0)(k_2 + k_3 + i0) \ldots (k_{n - 1} + k_n + i0)}
\\ [12pt]
\exp \left( 2i \left((k_1 + \ldots + k_n)x + (k_1^{2l - 1} + k_2^{2l -
1} + \ldots + k_n^{2l - 1})t_{(l)} \right)\right) \dst\frac{dk_1 dk_2 \ldots
dk_n}{(2
\pi i)^n}
\end{array}
$$
are conserved currents, namely, there is an operator $j_{S,
\lambda}(p_1, l, x, t_{(l)})$ which is a $\circs \circs$ -ordered product of
$i_{S,
\lambda}(p, l, x, t_{(l)})$ and its derivatives, with $p = 1, 2,\ldots ,$
such that
$$
\frac{d}{dt_{(l)}} i_{S, \lambda}(p_1, l, x, t_{(l)}) = \frac{d}{dx} j_{S,
\lambda}(p_1, l, x, t_{(l)})
$$

\vskip18pt

The proof is essentially the same as in [7]:

\noindent For a fixed $l$   let us introduce the following  notations :
\vskip6pt

\noindent $i(p) := i_{S, \lambda} (p, l, x, t_{(l)}) ;$

\noindent $A(x) \sim B(x)$,  if $A(x) - B(x) = \dst\frac{d}{dx} \txt\circs
C(x)\circs$, where $C(x)$ is a $\circs \circs$ polynomial in $i_{S, \lambda}
(p, l, x, t_{(l)})$ , $p = 1,2, \ldots$, and their derivatives;
\vskip18pt

The $\circs \circs$ product, defined in (5), has the properties
\vskip18pt
\noindent $\circs AB\circs = \circs BA\circs $,
\vskip12pt

\noindent $\circs(\circs AB\circs)C\circs = \circs A\circs BC\circs\circs$
\vskip12pt

\noindent $\dst\frac{d}{dx} \txt\circs AB\circs  = \txt\circs (\dst\frac{d}{dx}
A)B\txt\circs + \txt\circs A \dst\frac{d}{dx} B\txt\circs$,
\vskip18pt

therefore,from Lemma 2,
$$
\begin{array}{l}
\circs i(p_1) \dst\frac{d}{dx} i(p_2 + 1)\txt\circs +\txt\circs
(\dst\frac{d}{dx}i(p_1 + 1)) i(p_2) \txt\circs
\\ [12pt]
= -\dst\frac{1}{4} \frac{d}{dx}\left( \txt\circs i(p_1) \dst\frac{d^2}{dx^2}
i(p_2) +
i(p_2) \frac{d^2}{dx^2} i(p_1) - \frac{d}{dx} i(p_1) \frac{d}{dx}
i(p_2) + 4u \cdot i(p_1) i(p_2) \txt\circs \right) \sim 0
\end{array}
$$

{}From this relation , $\circs i(p_1) \dst\frac{d}{dx} i(p_2 + 1)\circs \sim
-\circs i(p_1+1) \dst\frac{d}{dx} i(p_2 ) \circs$ ; we also have
$\dst\frac{d}{dx} i(0) = 0 $ , and therefore,

$$
\circs i(p_1) \dst\frac{d}{dx} i(p_2 ) \circs \sim 0
$$

Taking the derivative of i(p) with respect to $ t_{(l)}$ and comparing with
Lemma 4, we get

$$
\dst\frac{d}{dt_{(l)}} i(p) \sim \txt\circs  i(p) \dst\frac{d}{dx} i(l
- 1) + i(p + 1) \frac{d}{dx} i(l - 2) + \ldots + i(p + l - 2) \frac{d}{dx} i(1)
\txt\circs \sim 0
$$

{\bf Fermionic case}
Let $\eta(k)$ be a grassman quantum field, and $\eta_S(k)$ - the
smeared grassman field.For grassman fields,we cannot just symmetryze the
product,because the result would be zero.But we can multiply polynomial
in $\eta$ and a function of $\{k\}$,in such a way that the product is
symmetric in $\{k\}$.For example,one can  substitute
$$
\begin{array}{l}
\left( O_S(k_1) O_S(k_2) \ldots O_S(k_n) \right)_W := \eta_S(k_1)
\eta_S(k_2) \ldots \eta_S(k_n)
\\ [8pt]
\cdot \det
\left|
\begin{array}{lllcl}
1& 1& 1& \ldots& 1\\
k_1& k_2& k_3& \ldots& k_n\\
k_1^2& k_2^2& k_3^2& \ldots& k_n^2\\
k_1^2& k_2^2& k_3^2& \ldots& k_n^2\\
\multicolumn{5}{c}{\ldots}\\
k_1^{n - 1}& k_2^{n - 1}& k_3^{n - 1}& \ldots& k_n^{n - 1}
\end{array}
\right|
\end{array}
$$
or,alternatively,$(O_S(k_1) \ldots O_S(k_n))_W = \epsilon_{i_1 \ldots
i_n}\eta_S(k_{i_1}) \ldots \eta_S(k_{i_n})$
in all the formulas, $n = 2, 3, \ldots$.

One has to do this change both in the definition of $\Psi_{S,
\lambda}$, $i_{S, \lambda}$ and in the definition of the $\circs \circs$
product.After such substitution,the conserved currents could be obtained by the
same construction as before.
In fact, in this construction $i_{S, \lambda}$ and $\Psi_{S, \lambda}$
is a sum of even and odd in $\eta$ terms.Therefore, for the conserved
quantities
we will have
$\dst\frac{d}{dt} i_{\scriptsize{\mbox{odd}}} = \dst\frac{d}{dx}
j_{\scriptsize{\mbox{odd}}}$ and $\dst\frac{d}{dt}
i_{\scriptsize{\mbox{even}}} = \dst\frac{d}{dx}
j_{\scriptsize{\mbox{even}}}$.

\section*{Conclusion.}
Starting from the basic field $O_S(k)$, we defined the observable
fields $u_{S, \lambda} (x,l, t_{(l)})$, $i_{S, \lambda} (p,l, x, t_{(l)})$,
and the regularized product $\circs \circs$. It is rather interesting
to notice,that , say,for $l = 2$, we have the following identity:
$$
-4 \cdot \dst\frac{\partial}{\partial t} u -
\dst\frac{\partial^3}{\partial x^3} u = 3 \frac{\partial}{\partial x}
\txt\circs u^2\txt\circs (x, t),
$$
where $\circs u^2\circs (x, t)$ is a new
observable field, obtained as a $\circs \circs$ product $\circs u u\circs$
If u(x,t) were a function,and with standart multiplication of functions,rather
than the $\circs \circs$ one, this would be the KdV equation.In the quantum
theory, we need some regularization to multiply operators at the same
point. With our regularization, we have the " Quantum KdV "relation.

We also have, with our definitions of $u$, $i$, and $\circs \circs$, that
$$
i_{S, \lambda} (2, x) = -\frac{1}{16} \txt\circs (3u^2 + u'')\circs (x)
$$
$$
i_{S, \lambda} (3, x) = \frac{1}{64} \txt\circs (10 u^3 + 10u u'' + 5(u')^2 +
u^{(IV)})\circs (x)
$$
$$
\ldots \quad \quad .
$$

\newpage
{\bf Remark.}
Notice,that in the noncommutative case,there is an alternative way to get
the KdV equation. Let $ O(k) $ be an element of a free algebra (which means,
that we didn't impose any commutation relations).Consider the formal series
$$
\begin{array}{l}
{\check u}( x, t) = 4\dst\sum^{\infty}_{n = 1}
\int  O(k_1) O(k_2) \ldots O(k_n)
\\ [12pt]
\dst\frac{k_1 + k_2 + \ldots + k_n}
{(k_1 + k_2 + i0)(k_2 + k_3 + i0) \ldots (k_{n - 1} + k_n + i0)}
\\ [12pt]
\exp \left( 2i \left((k_1 + \ldots + k_n)x + (k_1^{3} + k_2^{3}
 + \ldots + k_n^{3})t \right)\right) \dst\frac{dk_1 dk_2 \ldots dk_n}{(2
\pi i)^n}
\end{array}
$$
\centerline{(no symmetrization in the product)}
\vspace*{6pt}

Then, we have the following relation for the formal series :

$$
-4 \cdot \dst\frac{\partial}{\partial t} {\check u} -
\dst\frac{\partial^3}{\partial x^3} {\check u} = 3 \frac{\partial}{\partial x}
 ({\check u} {\check u} ),
$$

\centerline{(here the product is in the algebra)}

\vspace*{12pt}
\section*{\bf Discussion}.

\noindent To prove lemmas 1--5 we need only the identities for symmetric
functions in $n$ variables $k_1, \ldots, k_n$, $n = 1, 2, \ldots$.
The properties of the basic quantum field $O(k)$ are ,essentially,  not
important (except for the convergence of the series). We need only that the
products of smeared (regularized) basic field make sense and not singular on
the diagonals.

We postponed taking the Borel transform (4), and therefore our construction
depend on parameters, $\lambda$, which is,in fact the coupling constant,
and $S= (s_1, s_2)$ (``the cutoffs''). Therefore,if the operator $O(k)$ is
known,one might be able to study a``renormalization flow'' $\lambda(S)$. At the
``fixed points of the
flow'' $\lambda(S) O_S(k)$ is dimensionless, and the operators $i(p,
x), p = 1, 2, \ldots$ are of (formal) dimension $2p$.However,we do not know
the exact relationship between the fields $i(p, x)$ and operators in
conformal field theory ``at the fixed point.''

We feel that our construction is quite general, and similar approach will work
for other nonlinear models. However,we, in fact, didn't finish the
quantization.
For the quantization to be complete, one have to tell, what is the basic field
$O$ as operator, and what regularization to choose.

\section*{Acknowledgments}
	This paper appeared as a part of the work with prof. I.Gelfand in integrable
equations and their quantization.
	I would like to thank S. Lukyanov, A. B. Zamolodchikov,
A.S. Wightman , A.S. Fokas for fruitful discussions, and the High Energy Theory
group of Rutgers University for the support. The author is specially grateful
to A.B. Zamolodchikov for the suggestion that the quantum Sturm-Liouville
equation might exist.


\newpage

\begin{thebibliography}{9}

\bibitem{1}
E.K. Sklyanin, L.D.Faddeev," {\it Dokl. Akad. Nauk SSSR},{\bf 243} , 1430,
(1978)

\bibitem{2}
Faddeev,L.D., Takhtajan,L.A. : Hamiltonian Method in the Theory of Solitons.
NY:Springer 1987

\bibitem{3}
A.B.Zamolodchikov, {\it Adv. Stud. Pure Math.}, {\bf 19},641-674,1989.

\bibitem{4}
B.Feigin, E.Frenkel, Integrals of Motion and Quantum Groups, preprint
hep-th/9310022 .

\bibitem{5}
B.A. Kupershmidt, P.Mathew., {\it Phys.Lett. B} {\bf227}, 245-249, 1989.

\bibitem{6}
V. Bazhanov, S. Lukyanov, A.B. Zamolodchikov, "Integrable Structure of
Conformal Field Theory, Quantum KdV Theory and Thermodynamic Bethe Ansats,
preprint hep-th/9412229, and references therein.

\bibitem{7}
I.M. Gelfand , L. A. Dikii,
{\it Russ.\ Math.\ Surv.} 30 (5) ,77--113, 1975.

\bibitem{8}
I.M. Gelfand, A.S. Fokas and M.V. Zyskin, "Nonlinear Integrable Equations
and Nonlinear Fourier Transform", hep-th/9504042 .

\bibitem{9}
R.F. Streater, A.S. Wightman, PCT, Spin and Statistics, and all that.
Addison-Wesley, 1989.

\bibitem{10}
J. Glimm, A. Jaffe. Quantum Physics. Springer- Verlag, 1987.

\bibitem{11}
I.M. Gelfand, G.E. Shilov. Generalized Functions. Academic Press, 1964.

\end{thebibliography}
\end{document}